\documentclass[a4paper]{article}
\usepackage{amsmath}

\usepackage[hmarginratio=1:1,top=40mm,columnsep=20pt]{geometry} 

\usepackage{titlesec} 
\titleformat{\section}[block]{\large\scshape\centering{\Roman{section}.}}{}{1em}{} 

\begin{document}

\def\grad{{{\rm grad}\; }}
\def\div{{{\rm div}\; }}
\def\dV{{\;d^3 {\bf r}}}
\def\D{{{\bf {D}} }}
\def\r{{{\bf {r}} }}
\def\r{{{\bf r} }}
\def\ss{{\rho_f}}
\newcommand{\E}{{\mathbf{E}}}
\def\q{{\bf q}}
\def\P{{\bf P}}
\def\ash{{{\rm asinh} }}

\date{}
\title{\vspace{-15mm}\fontsize{20pt}{10pt}\selectfont\textbf{Legendre transforms for electrostatic energies\label{pb}}} 

\author{
\large
\noindent
\textsc{Justine S. Pujos and A.C. Maggs}\\ 
\normalsize CNRS Gulliver, ESPCI, 10 rue Vauquelin, 75231 Paris, Cedex 05. \\
\vspace{-5mm}
}

\maketitle

\begin{abstract}
  We review the use of Legendre transforms in the formulation of
  electrostatic energies in condensed matter. We show how to render
  standard functionals expressed in terms of the electrostatic
  potential, $\phi$, convex -- at the cost of expressing them in terms
  of the vector field $\D$. This leads to great simplification in the
  formulation of numerical minimisation of electrostatic energies
  coupled to other physical degrees of freedom. We also demonstrate
  the equivalence of recent functionals for dielectrics derived using
  field theory methods to classical formulations in terms of the
  electric polarisation.
\end{abstract}

\section{Introduction}

The Legendre transform is a powerful tool with multiple applications
in physics \cite{zia}. In classical mechanics it allows one to
interchange the Lagrangian and Hamiltonian viewpoints; in
thermodynamics one regularly transforms ensembles to simplify
calculations, choosing the ensemble which most closely idealizes a
given experimental setup. In this article we will demonstrate the
utility of Legendre transforms in reformulating energies and free
energies in electrostatics, with a particular eye for numerical
applications and mean field theory.

Our principal motivation for a deeper study of this transformation
applied to electrostatic problems is quite practical.  Many
formulations of (free) energy functions in condensed matter physics
involve the electrostatic potential, an important example is the
Poisson-Boltzmann energy functional in the theory of ionic
solutions. However when we examine closely these functionals we see
that they are concave functions of the potential. While the stationary
value of the functional is indeed the correct value of the energy that
we wish to study, the concavity leads to complications in many
situations. In particular we can not perform a simultaneous
minimisation of both electrostatic and configurational energies in a
simulation. One is generally obliged to fully solve by iterative
methods the electrostatic problem at each time step of the iteration
over the configurational degrees of freedom - such as densities or
polymer configuration. This leads to codes which are complicated to
write, and sometimes slow to run.

We remind the reader that for a convex function $f(x)$, its Legendre
transform is defined \cite{zia} from the expression
\begin{equation}
  \mathcal{L} \left[f \right] \left( s \right) = g(s) = s  x - f(x)
  \label{eq:legendre}
\end{equation} 
where on the right of eq.~(\ref{eq:legendre}) we express $x$ as a
function of $s$ from the equation $s=\frac{df}{dx}$.  This
transformation is an involution: $\mathcal{L} \left[ g \right] \left(
  x \right) = f(x)$.  The simplest example is an Hookian spring for
which $f(x) = kx^2/2$ ; then the transformation of
eq.~(\ref{eq:legendre}) gives $g(s) = s^2/(2k)$. We will also use the
notation $\mathcal{L}(f) = \tilde f$.  In this article we show that by
introducing new variational parameters in a free energy with the help
of Lagrange multipliers and then performing a Legendre transform of
the resulting free energy we can find functionals that are convex in
all degrees of freedom; we will illustrate this with a mean field
formulation of phase separation coupled to electrostatic interactions.

We now illustrate the approach with the simplest possible
electrostatic problem, interaction between free charges, $\rho_f$ in a heterogeneous
dielectric medium: Consider the energy functional expressed in terms
of the electric potential $\phi$.
\begin{equation}
  U= \int \left \{ -\frac{ \epsilon(\r) (\nabla \phi)^2}{2} + \rho_f \label{eq:phien}
    \phi\right \} \dV
\end{equation}
The variational equation for the field is then the Poisson equation:
\begin{equation}
  \div \epsilon(\r) \grad \phi = - \rho_f(\r)
\end{equation}
Substituting the solution of the Poisson equation in the electrostatic
energy we find
\begin{equation}
  U = \frac{1}{2} \int \phi(\r) \rho_f(\r) \dV
\end{equation}
We convert the variational problem for the potential by introducing
the new variable $\E= - \nabla \phi$. To do so we introduce the
(vector) Lagrange multiplier $\D$. The stationary point of  eq.~(\ref{eq:phien}) is
identical to that of  the following expression \cite{courant}:
\begin{equation}
  U = \int \left \{ - \frac{\epsilon(\r) \E^2 }{2} + \rho_f \phi 
    + \D \cdot (\E+ \nabla \phi) \right \} \dV \label{eq:full}
\end{equation}
It is at this point that we recognise that the variational equations
for $\E$ correspond to a Legendre transform with dual variable
$\D$. We also integrate by parts the product $\D\cdot \nabla \phi$ to
find $-\phi\, \div \D$, dropping boundary terms assumed to be
zero. Thus the stationary point of eq.~(\ref{eq:phien}) is identical to
the stationary point of 
\begin{equation}
  U = \int \left \{ \frac{\D^2}{2 \epsilon(\r)} + \phi(\rho - \div \D )
  \right \} \dV
\end{equation}
Variations in $\phi$ now impose Gauss' law, $\div\D-\rho_f=0$, while
the energy has been rendered convex by the transformations introduced,
\cite{localpb, aux}.

We now illustrate applications of this transformation to two
problems: Firstly the theory of phase separation of immiscible fluids
in the presence of electrostatic interactions due to ions.  Secondly we
provide a translation between two very different visions of the theory of
dielectrics. Recent formulations of implicit dielectrics pass by
elaborate field-theory mappings and find a generalised
Poisson-Boltzmann equation with a Langevin correction. We show how to
map this description onto a free energy expressed in terms of a
polarisation field with long-ranged dipolar interactions. We believe
that these equivalent descriptions can lead to a deeper understanding
of the underlying physics.

In the following we will work with free energy densities, rather than
the integrated energies and we (silently) integrate by parts when
needed.

\section{Phase separation coupled to electrostatics}

A mixture of two solvents (A and B) near their miscibility limit and in
the presence of salt displays interesting properties which have been
explored in recent experiments \cite{bonn}. Density fluctuations
couple to the dielectric properties of the medium, and in turn
influence the partition of ions in the fluctuating solvent field.  The
experimental system has turned out to be very rich, and allows one to
adjust the effective interaction between colloidal particles using
temperature as a control parameter.

A simplified theoretical description of such systems is given in
\cite{D1,onuki} who propose the following free energy density
expressed in terms of the densities and potentials:
\begin{align} 
  f(\phi ,& \Psi , c_+, c_-)= {f_m}(\Psi) - \frac{1}{2}
  \epsilon(\Psi) (\nabla\phi)^2
  + (c_+ - c_-)e\phi \nonumber  \\
  & - \left( \Delta u^+ c_+ + \Delta u^- c_- \right)\Psi + k_B T \sum_j
  ( c_j \ln{(c_j/c_{j0}) } - c_j)
  \label{eq:Ts}
\end{align}
$\Psi$ describes the composition fluctuations of the fluid
mixture. $c_+$ and $c_-$ are the concentration of positive and
negative monovalent ions, with $ \Delta u^+$ and $\Delta u^-$ their
relative preferences between a A-liquid environment and a B-liquid
environment. As above $\phi$ is the electrostatic potential. We see
that fluctuations in concentration couple via $\epsilon(\Psi)$
to a coupling with the concentration fluctuations of the ions.

${f_m}(\Psi)$ includes all the terms that are only dependent on
$\Psi$: ${f_m}(\Psi) = f_0(\Psi) + \frac{c}{2} (\nabla\Psi)^2 -\mu
\Psi$; with $f_0(\Psi)$ the free energy due to the mixing of the two
solvents.  It can, for exemple, be written as a binary mixture free
energy density: $f_0(\Psi) \propto \Psi \log (\Psi) + (1-\Psi)
\log(1-\Psi) + \chi \Psi (1-\Psi)$, with $\chi$ the Flory parameter
\cite{D2,yoav}, or as a Landau expansion $f_0(\Psi) \propto \alpha \left(\Psi-\Psi_c \right)^2 +
\gamma \left(\Psi-\Psi_c \right)^4 $, with $\alpha$ being temperature dependent, 
$\gamma$ positive, and $\Phi_c$ the critical composition \cite{D1}.

Optimising eq.~(\ref{eq:Ts}) over $c_+$ and $c_-$, the density becomes
:
\begin{align}
  f(\Psi,\phi) = {f_m}(\Psi) -& \frac{1}{2} \epsilon(\Psi)
  (\nabla\phi)^2 - k_B T c_{0+} \exp(\beta \Delta u^+ \Psi - \beta e
  \phi)
  \nonumber \\
  -& k_B T c_{0-} \exp(\beta \Delta u^- \Psi + \beta e \phi)
  \label{eq:Tsori2}
\end{align}
With a symmetric electrolyte: $c_{0+}=c_{0-}$ , and if we assume the ions are
similar in their interaction with the solvents : $\Delta u^+ = \Delta
u^-$, $f(\Psi,\phi)$ simplifies into :
\begin{equation}
  f(\Psi,\phi)= {f_m}(\Psi) - \frac{1}{2} \epsilon(\Psi) (\nabla\phi)^2
  - 2 k_B T c_{0} \exp(\beta \Delta u \Psi) \cosh( \beta e \phi)
  \label{eq:Tsori2bis}
\end{equation}
We recognise here a generalisation of the well known Poisson-Boltzmann
functional for a symmetric electrolyte.  The description is adapted to
analytical solutions but in the monophase region of the phase diagram
$f(\Psi,\phi)$ is convex in $\Psi$ but concave in $\phi$. In
complicated geometries if one wishes to minimize this free energy
numerically one has to solve saddle point equations, simple
minimization will not give the correct answer.  We now implement the
transformation introduced above from the potential $\phi$ to the electric displacement
$\D$ and use the fact that the Legendre transform of $\cosh$ is :
\begin{equation}
  \begin{aligned}
    \mathcal{L}[A \cosh (B \phi) ](\xi) &=
    A \left[ {\xi}/(AB) \ash \left({\xi}/{AB} \right) - \sqrt{\left({\xi}/{AB} \right)^2+1}  \right] \\
    = & A g \left( {\xi}/{AB} \right )
  \end{aligned} \label{eq:TlegCH}
\end{equation}
After some calculation we find
\begin{equation}
  \begin{aligned}
    f(\Psi,\D)= {f_m}(\Psi) + \frac{\D^2}{2\epsilon(\Psi)} + 2
    k_B T c_{0} e^{\beta \Delta u \Psi} g \left( \frac{\div(\D)
        e^{-\beta \Delta u \Psi}}{2 c_0 e} \right)
  \end{aligned} \label{eq:Tsori2tris}
\end{equation} 
We have thus reached our objective: we have built an equivalent
description of the system with the stationary conditions conserved and
a local and convex function. The disadvantage is that there are more
degrees of freedom in the vector field $\D$ than in the scalar field
$\phi$, but the advantage is that a global minimising principle can 
be used and the functional can be directly programmed for the
solution of the coupled electrostatic-phase separation problem.

We note that mean field description of the packing of DNA in a virus
\cite{rudi} contains many similar theoretical features and is also
amenable to similar transformations. In this problem the field $\Psi$
corresponds to the square root of the monomer density.

\section{From Poisson-Langevin to Polarization}
We now consider theories of explicit Langevin dipoles and how these
can be incorporated into the formulation of the free energy in terms
of convex free energy functions. Recent work on improving the
description of solvation of proteins \cite{marc} has considered an
explicit model for the solvent in terms of Langevin dipoles. If we
neglect the volume of ions and dipoles they find the free energy
density for a mixture of symmetric ions and neutral dipoles :
\begin{equation}
  f = \rho_f \phi- \frac{\epsilon_0 (\nabla \phi)^2}{2} - 2 \lambda_{ion}
  \cosh{(\beta q \phi)} -  \lambda_{dip} \frac{\sinh(\beta p_0 | \nabla
    \phi| ) }{\beta p_0 |\nabla \phi|} \label{eq:langevin}
\end{equation}
where to simplify the presentation we have neglected effects of finite
ion size. The parameters $\lambda$ are related to the chemical
activities of the ions and the dipoles.

As in previous work \cite{pbrecent} we start by using a Lagrangian
multiplier, $\D$ to replace ($\nabla \phi)$ by its electrostatic
equivalent $-\E$. We find
\begin{equation}
  f= \rho_f \phi - \frac{\epsilon_0  \E^2}{2} - g(\phi) -  h({\E})
  + \D\cdot (\nabla \phi +\E) \label{eq:E}
\end{equation}
where $h(\E)$ is the free energy density due to the dipoles and
$g(\phi)$ the free energy due to free ions.  We now diverge from our
previous treatment and introduce a new variable $\P$ which we will
show is the physical polarisation variable. We do this by performing a
Legendre transform on $h(\E)$ to find $\tilde h (\P)$; we then find
\begin{equation}
  f = \phi(\rho_f -\div \D) - \frac{ \epsilon_0 \E^2 }{2} - g(\phi) + 
  \tilde h({\bf P})  + \E \cdot ( \D -  {\bf P} ) \label{eq:P}
\end{equation}
Clearly by definition of the Legendre transform performing variations
with respect to $\P$ on eq.~(\ref{eq:P}) gives eq.~(\ref{eq:E}).

We now perform two more transforms firstly to eliminate the potential,
but secondly to eliminate the electric field $\E$. We find
\begin{equation}
  f = \frac{ (\D - {\bf P})^2 } {2 \epsilon_0 } +\tilde h({\bf P}) + \tilde g
  (\rho_f - \div \D)
  \label{ralf}
\end{equation}
In the absence of free ions the function $\tilde g$ reduces to the
constraint of Gauss' law. This is exactly the form postulated in
\cite{everaers}. It is particularly transparent for understanding the
physical limits on response functions \cite{dolgov, kornyshev} and the
origin of the negative dielectric constant observed in structured
fluids.

We will now work from eq.~({\ref{ralf}}) to demonstrate its
equivalence to other formulations of electrostatic interactions
expressed in terms of the polarization $\P$. To do this we will
eliminate the variable $\D$, which will bring us back to other more
familiar forms for the electrostatic energy at the cost of
re-introducing long-ranged dipole-dipole interactions between the
polarization variables.
\subsection*{Eliminating the Displacement field}
Let us work in the limit where linear response is valid, in which case
we expand $\tilde h$ to quadratic order:
\begin{equation}
  \tilde h(\P) = \frac{\P^2}{2 \epsilon_0 \chi}
\end{equation}
where $\chi$ is a material parameter. Taking variations of
eq.~(\ref{eq:P}) with respect to $\P$ and then $\E$ we find 
that
\begin{equation}
  \P = \epsilon_0 \chi \E, \quad \epsilon_0 \E = \D - \P
\label{eq:chi}
\end{equation}
Thus the parameter $\chi$ is the electric susceptibility of the
medium.  The polarisation variable is indeed playing the role we
expect from standard treatments of Maxwell's equations. The free
energy of the fluctuating dipoles (in the absence of free ions) can
then be found from the functional
\begin{equation}
  f=  \frac{(\D -\P)^2}{2 \epsilon_0 }
  + \frac{\P^2}{2 \epsilon_0 \chi(\r)}
  - \phi (\div \D - \rho_f)
 \label{eq:U}
\end{equation}
where the last term is a Lagrange multiplier for the constraint of
Gauss' law.  On taking variations of eq.~(\ref{eq:U}) with respect to
$\D$ we find that
\begin{equation}
\D - \P = - \epsilon_0 \nabla \phi
\end{equation}
Thus the free energy density can be written as
\begin{equation}
f= \epsilon_0 \frac{(\nabla \phi)^2}{2} + \frac{\P^2}{2 \epsilon_0 \chi} \label{eq:dens}
\end{equation}
where 
\begin{align}
\epsilon_0 \nabla^2 \phi &= - \rho_f + \div \P \nonumber \\
\phi(\r ) &= \int \frac{1}{4 \pi \epsilon_0 |\r-\r' |} \left ( \rho(\r') - \div \P(\r')  \right
)\dV'  \label{eq:phi}
\end{align}
We now substitute eq.~(\ref{eq:phi}) in eq.~(\ref{eq:dens}) and
introduce the bare electric field $\E_0$ as follows: $\E_0 = -\nabla
\phi_0$ with $\epsilon_0 \nabla^2 \phi_0 = -\rho_f$.  The free energy
density can then be expressed in terms of the polarization as
\begin{align}
  U= \frac{1}{2} \int & \frac{\div \P(\r) \div \P(\r')}{4 \pi
    \epsilon_0 |\r -\r' |} \dV \dV' +\nonumber \\ \int & \left \{
    \frac{\epsilon_0 \E_0^2}{2} -\E_0\cdot \P + \frac{\P^2}{2
      \epsilon_0 \chi(\r)} \right \} \dV
  \label{eq:marcus}
\end{align}
This formulation of the free energy is widely used in theoretical
chemistry and is that used by \cite{marcus}

The energy eq.~(\ref{eq:marcus}) can be expressed in an even more
physically transparent manner by integrating by parts the first double
integral to transfer the derivatives from the polarization to the
function $1/|\r-\r'|$. If we do this we find that the double integral
is transformed to
\begin{equation}
  \frac{1}{2}\int \P(\r) T(\r-\r') \P(\r') \dV \dV'
\end{equation}
where the dipole operator is
\begin{equation}
  T(\r) = \left [ \frac{1- 3|\r \rangle \langle\r | }{4 \pi r^3 \epsilon_0} +
    \frac{\delta(\r)}{3 \epsilon_0 } \right ] \label{eq:T}
\end{equation}

We now proceed in a more abstract manner, considering that the
polarisation variables are assembled into a vector and the dipolar interactions
form a matrix, $\bar T$. Eq.~(\ref{eq:marcus}) is simply a quadratic form in $\P$, so that
\begin{equation}
  U= \frac{ \P(\bar K+\bar T)\P }{2} -\E_0 \cdot \P + \epsilon_0 \frac{\E_0^2}{2}
\end{equation}
with the diagonal matrix
$\epsilon_0 K_{\r,\r}=\chi^{-1}_\r$. If we now calculate the
response of the polarization field to an external perturbation $\E_0$
we find
\begin{equation}
  \P = \frac{1}{\bar K+\bar T} \E_0
\end{equation}

The total electric field is then given by two contributions, the
original imposed field $\E_0$ and that due to the dipole density $\P$:
\begin{align}
  \E &=\E_0 - \bar T \P\\
  &= \frac{\bar K}{\bar K+\bar T} \E_0
\end{align}
We also use
\begin{equation}
  \D-\P = \epsilon_0 \E
\end{equation}
to find
\begin{equation}
  \D= \frac{I + \epsilon_0 \bar K}{\bar K+\bar T} \E_0
\end{equation}
Again all these equations are non-local - since they involve the
long-ranged operator $\bar T$ but we find that
\begin{equation}
  \D= (1+\chi_\r ) \epsilon_0 \E = \epsilon \E
\end{equation}
a purely local constitutive equation between the electric field and
the electric displacement.

We conclude that by careful study of the mean field equations coming
from the formulation of the dielectric properties of a medium in
eq.~(\ref{eq:langevin}) we have been able to derive the equivalence to
the standard continuum formulation of electrostatic arising from
Maxwell's equations.
\section{Conclusions}

We have shown that the Legendre transform can be used to translate
between multiple forms of the energy in mean-field theories. All the
formulations are numerically equivalent but different forms put the
emphasis on different degrees of freedom in electromagnetism. For
numerical work it is advantageous to work with a formulation which is
both convex and local. This is achieved in Poisson-Boltzmann theory by
choosing the electric displacement $\D$ as the fundamental
thermodynamic field. In this way all physical degrees of freedom can
be treated in an equivalent manner in numerical solvers. It is no
longer necessary to completely solve the electrostatic problem for
each iteration of other external degrees of freedom. Very similar
conclusions have also been found in  quantum chemistry \cite{quantumDP}

We have also demonstrated that energy functionals for dielectrics can
be translated into equivalent forms by introducing the physical
polarization. We then mapped a linearised form of the theory to the
Marcus energy function, widely used in the theoretical chemistry literature.


\end{document}